%% file: AFB_PHG_DPF_2015-2.tex
\documentclass[12pt]{article}
\usepackage{graphicx}
\usepackage{float}  

\input econfmacros.tex    

\newcommand\pubnumber{DPF2015-414}
\newcommand\preprint{FERMILAB-CONF-15-395-E}
\newcommand\pubdate{September 25, 2015}

\def\D0-Fermilab{On behalf of the D0 Collaboration\\
Fermi National Accelerator Laboratory\\ Batavia, Illinois, 60510, USA}
\def\support{\footnote{Work supported by: xxx do I really need this Ask Heath OConnell or Kathryn Duerr}}

\def\Title#1{\begin{center} {\Large #1 } \end{center}}
\def\Author#1{\begin{center}{ \sc #1} \end{center}}
\def\Address#1{\begin{center}{ \it #1} \end{center}}

\newcommand\pubblock{\rightline{\begin{tabular}{l} \pubnumber\\
         \preprint\\ \pubdate  \end{tabular}}}
\newenvironment{Abstract}{\begin{quotation}  }{\end{quotation}}
\newenvironment{Presented}{\begin{quotation} \begin{center} 
             PRESENTED AT\end{center}\bigskip 
      \begin{center}\begin{large}}{\end{large}\end{center} \end{quotation}}
\def\Acknowledgments{\bigskip  \bigskip \begin{center} \begin{large}
             \bf ACKNOWLEDGMENTS \end{large}\end{center}}
\def\D0-Fermilab{On behalf of the D0 Collaboration\\
Fermi National Accelerator Laboratory\\ Batavia, Illinois, 60510, USA}
\def\support{\footnote{Work supported by: Fermi Research Alliance, LLC under Contract No. DE-AC02-07CH11359
with the United States Department of Energy.}}

\RequirePackage{lineno}

\begin{document}


\begin{titlepage}
\pubblock

\vfill
\Title{Recent D0 Measurements of Forward-Backward Asymmetries for $p\overline{p} \rightarrow B^\pm, \Lambda_b^0$, and $\Lambda_s^0$ Production}  
\vfill
\Author{ Peter H. Garbincius\support}
\Address{\D0-Fermilab}
\vfill
\begin{Abstract}
The Forward-Backward Asymmetries in the production of $B^\pm, \Lambda_b^0$, and $\Lambda_s^0$ particles for rapidities $\mid y \mid$  $<$ 2 are measured by D0 for $p\overline{p}$ 
collisions at $\sqrt{s}$ = 1.96 TeV at the Fermilab Tevatron. $A_{FB}(B^\pm)$ and $A_{FB}(\Lambda_b^0)$ are consistent with zero, while $A_{FB}(\Lambda_s^0)$ exhibits a statistically significant rise with increasing $\mid y \mid$.
\end{Abstract}
\vfill
\begin{Presented}
DPF 2015\\
The Meeting of the American Physical Society\\
Division of Particles and Fields\\
Ann Arbor, Michigan, August 4--8, 2015\\
\end{Presented}
\vfill
\end{titlepage}
\def\thefootnote{\fnsymbol{footnote}}
\setcounter{footnote}{0}

\section{Introduction and Motivation} 


Alternative versions of this topic might be, ``do secondary particles retain some `memory' of the direction of the primary parent particles?''  Of course they do in the beam fragmentation (large Feynman-$x$) regions as studied in Fixed Target and ISR experiments.  But what about for the central regions near rapidity \cite{rapidity} $y$ = 0 for hadron colliders? What mechanisms are at play for the forward-backward asymmetries, denoted $A_{FB}$?  What occurs in the transition between the central and beam fragmentation regions?

The definition of Forward and Backward is usually pretty intuitive. Forward particles are those where the produced quarks (anti-quarks) or charge follows the direction of the beam particles with similar quarks, anti-quarks, or charge.  The proton direction ($y > 0$) is forward for $t$-quarks (and jets), $b$-quarks (and jets or decay particles), or baryons ($\Lambda_b^0$ and $\Lambda_s^0$)  for these studies), while the antiproton direction ($y < 0$) is forward for $\overline{t}$-quarks (and jets), $\overline{b}$-quarks (and jets or decay particles), or antibaryons.  Forward $B^-$ mesons ($b\overline{u}$) follow the proton direction, while forward $B^+$ ($\overline{b}u$) mesons follow the antiproton direction.  At the $pp$ LHC, all baryons would be considered ``forward'' and antibaryons ``backward'' regardless of their rapidity.

The forward-background asymmetries for the production of $t\overline{t}$ \cite{CDF_ttbar} \cite{D0_ttbar} and $W^\pm$ 
\cite{D0_Wasym} have been extensively studied for $p\overline{p}$ collisions at the Fermilab Tevatron.  Unlike the symmetric $pp$ initial state at the LHC, the $p\overline{p}$ state has the beam valence quarks traveling in the initial $p$ direction ($y > 0$)  and the beam valence antiquarks traveling in the initial $\overline{p}$ direction ($y < 0$).  Is this forward-backward asymmetry in the direction of motion of the initial state quark charges retained for the produced particles?

\begin{figure}[H]
\centering
\includegraphics[height=1.75in]{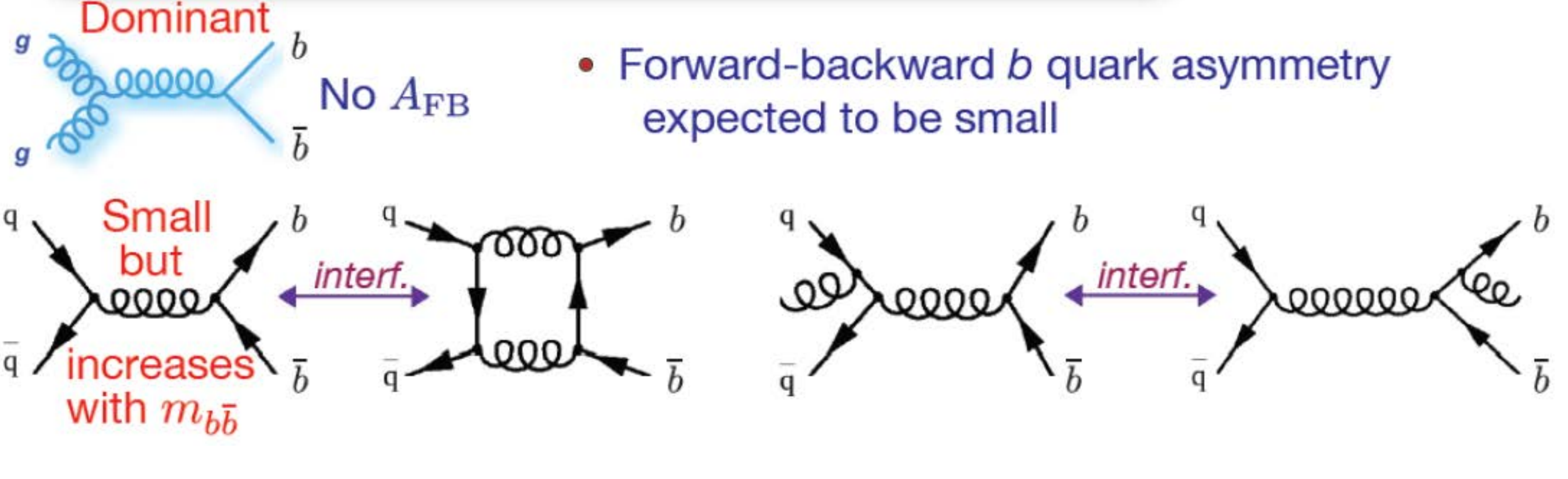}
\caption{Graphs for $p\overline{p}$ production of $b\overline{b}$ pairs.  For $p\overline{p} \rightarrow t\overline{t}$ pairs, the $q\overline{q}$ annihilation graphs dominate.}
\label{fig:graphs}
\end{figure}

Figure~\ref{fig:graphs} shows the first order $p\overline{p}$ production graphs for the central, small $\mid y \mid$, region for gluon fusion (dominant for $b\overline{b}$) and $q\overline{q}$ annihilation (dominant for $t\overline{t}$).  Neither of these first order graphs produce a forward-backward asymmetry.  However, interference with the loop and initial and final state gluon radiation graphs can produce a significant forward-backward asymmetry.   
Recent measurements indicate that $A_{FB}(t\overline{t})$ $\approx$ + 13\%, significantly different from zero.  
These experiments have motivated the calculation of the Next-to-Leading Order (NLO) and Next-to-Next-to-Leading Order (NNLO) processes to attain agreement with the observations.

For $b\overline{b}$ production in the central region, the gluon fusion graph is dominant and the higher order interference terms are small, leading to an expectation that $A_{FB}(b\overline{b})$ would be small.

The Fermilab Tevatron is a good place to study $A_{FB}$ since the symmetry of the initial $p\overline{p}$ state has net charge $C$ = 0 and net baryon number $B$ = 0.  The production cross sections for particles and antiparticles are expected to be related as

		\[ \sigma(particle, y) = \sigma(\overline{particle}, -y) \]

compared to the LHC where the initial $pp$  state has $C$ = +2, $B$ = +2, and 

		\[ \sigma(particle, y) = \sigma(particle,-y) \]

and where	$\sigma(particle, y)$ is not necessarily equal to $\sigma(\overline{particle}, y)$.
Thus the $p\overline{p}$ Tevatron allows complementary set of forward-backward comparisons with a complementary set of expected symmetries for the produced particles.


\section{The D0 Detector \cite{D0_detector} } 

\begin{figure}[H]
\centering
\includegraphics[height=2.67in]{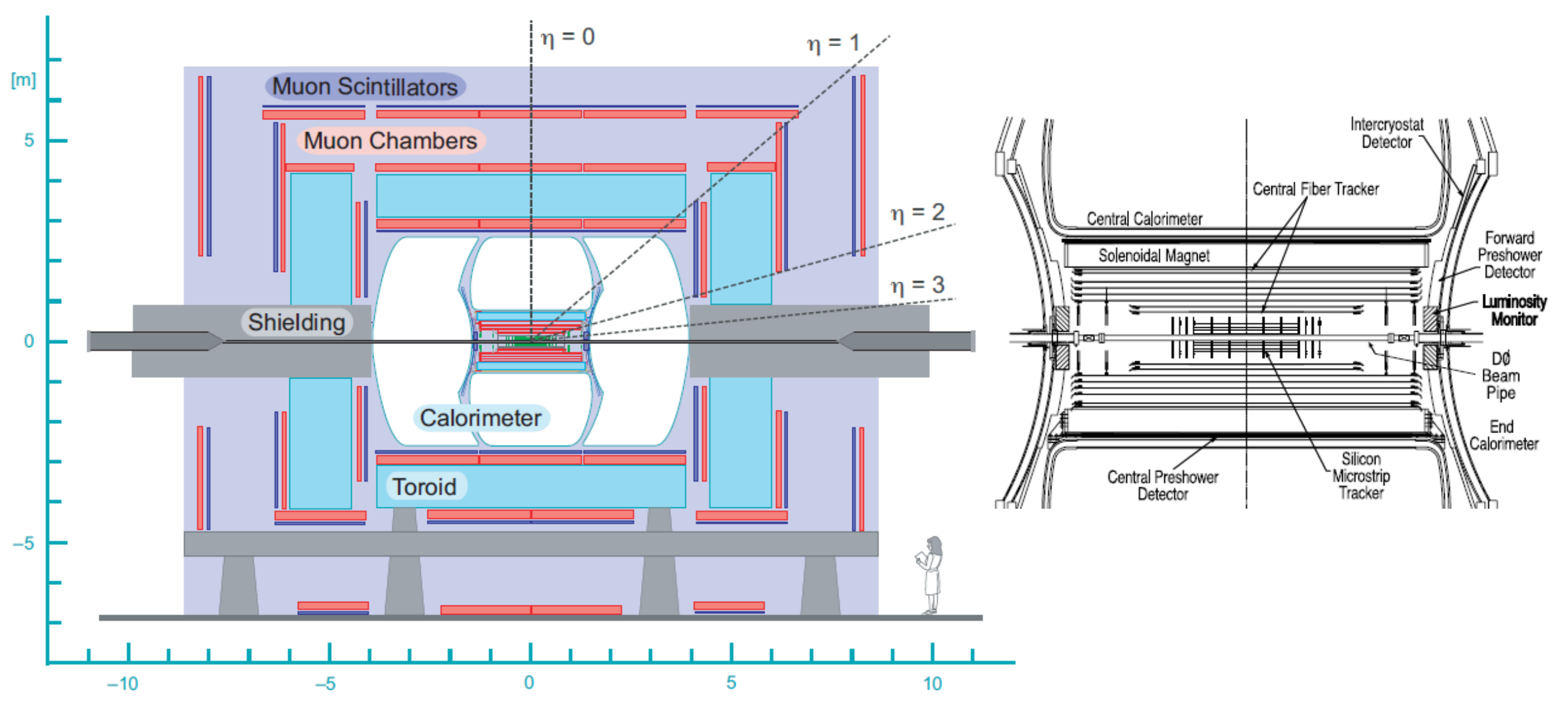}
\caption{The D0 Experiment for Run II of the Fermilab Tevatron \cite{D0_detector}.}
\label{fig:D0_experiment}
\end{figure}

The D0 experiment, illustrated in Figure~\ref{fig:D0_experiment}, is, at least to first order, forward backward symmetric with respect to rapidity $\pm$ $y$ or pseudorapidity \cite{rapidity} $\pm$ $\eta$.

Since the charged particle tracking is based on scintillating fibers and silicon strips whose performance is independent of the magnetic field direction, D0 regularly reverses the polarities of the tracking solenoid and muon toroid magnets which cancels many of the detector-related or detector-induced symmetries in the raw data.

The D0 results presented here are based on the full Tevatron Run II data set of 10.4 $fb^{-1}$ integrated luminosity for $p\overline{p}$ collisions at $\sqrt{s}$ = 1.96 TeV.

\section {Analysis for $A_{FB}$}  

Prior Tevatron and LHC studies of the asymmetries for the $t\overline{t}$ and $b\overline{b}$ systems used jets tagged by the charge of the decay leptons, and used  as their metric the difference in rapidities 
$\Delta y = y_{q-jet}-y_{\overline{q}-jet}$.  
In this analysis, D0 studies asymmetries for the single particle 
$B^\pm$, $\Lambda_b^0/\overline{\Lambda}_b^0$, and $\Lambda_s^0/\overline{\Lambda}_s^0$ inclusive production, without requiring the detection of both members of the 
$B^-$ and $B^+$, or $\Lambda_b^0$ and $\overline{\Lambda}_b^0$, or $\Lambda_s^0$ and $\overline{\Lambda}_s^0$ pair in the same event.

The Forward-Backward Asymmetry for the production of a particle at rapidity $y$ is defined as
	
\[ A_{FB}(y) = \frac{\sigma_F(y) - \sigma_B(y)}{\sigma_F(y)+\sigma_B(y)} = \frac{DIFF}{SUM} \]

where $\sigma_F(y)$ ($\sigma_B(y)$) is the differential cross section $d\sigma_F(y)/dy$ ($d\sigma_B(y)/dy)$) for the Forward (Backward) production process.  The asymmetry, as usually defined, is the difference divided by the sum for processes being compared.

We measure the $A_{FB}$ asymmetry simultaneously for the particle and antiparticle and $y < 0$ and $y > 0$ hemispheres and define

\footnotesize
\[
A_{FB}(\mid y \mid) = \frac{\sigma_F(particle,y>0)+\sigma_F(\overline{particle},y<0) - \sigma_B(particle,y<0) - \sigma_B(\overline{particle},y>0)}{
            \sigma_F(particle,y>0)+\sigma_F(\overline{particle},y<0)+\sigma_B(particle,y<0)+\sigma_B(\overline{particle},y>0)}.
\]
\normalsize

To assure no mixing of the forward and backward hemispheres due to reconstruction errors, particles with $\mid y \mid$ $<$ 0.1 or $\mid \eta \mid$ $<$  0.1 are excluded from these sums.  This assures that 99.9\% of the $B^\pm$ have the same sign of $y$ or $\eta$ of its parent quark.  For reference, the distribution of ($\eta_{b-quark} - \eta_{B-meson}$) has an r.m.s. width of 0.11 unit.

The range of these $A_{FB}$ measurements is 0.1 $<$ $\mid y \mid$ $<$ 2 units of rapidity.

\section{$A_{FB}(B^{\pm})$ \cite{AFB_B}}  


Charged $B$-mesons were reconstructed as $B^\pm \rightarrow J/\psi K^\pm$ with subsequent decay $J/\psi \rightarrow \mu^+ \mu^-$.   Since D0 does not have identification of hadronic particles, the charged decay product was assigned the mass of the kaon.  The muons were each required to have transverse momentum $p_T(\mu) >$ 1.5 GeV and $p_T(K) >$ 0.7 GeV.   A Boosted Decision Tree (BDT) selection was used to help separate signal from background. 

\begin{figure}[H]
\centering
\includegraphics[height=3.15in]{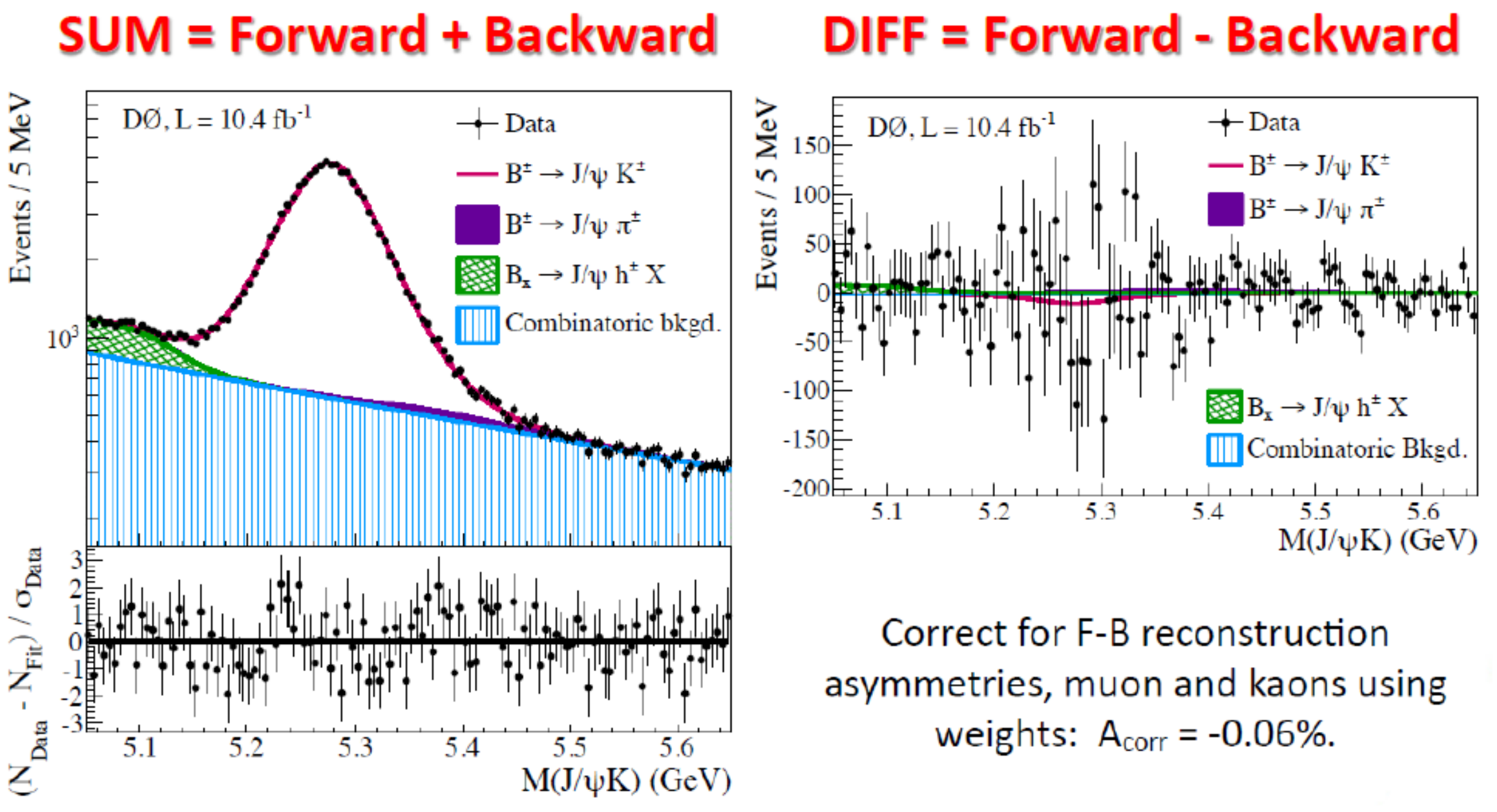}
\caption{Sum (Forward + Backward) and Difference (Forward \(-\) Backward) Mass distributions for $B^\pm$.}
\label{fig:AFB_B}
\end{figure}

An unbinned likelihood distribution was used to fit the invariant mass distributions for the 
$SUM = Forward+Backward$ 
and $DIFF = Forward-Backward$ candidates including four contributions:
1)	An exponential combinatorial background  distribution;
2)	A double Gaussian for the $B^\pm \rightarrow J/\psi K^\pm$ state of interest;
3)	A double Gaussian for incorrectly  identified $B^\pm \rightarrow J/\psi \pi^\pm$; and
4)	An arctangent threshold function for partially reconstructed $B_x \rightarrow J/\psi h^\pm X$.
For illustrative purposes, the Sum and Differences are plotted as binned distributions in Figure~\ref{fig:AFB_B}.  There are 89,000 $B^\pm \rightarrow J/\psi K^\pm$ candidates in the $SUM$ distribution.  $A_{FB}$ is the number of events in red $DIFF$ distribution, divided by the number of events in the red $SUM$ distribution for $B^\pm \rightarrow J/\psi K^\pm$. 

Integrated over 0.1 $<$ $\mid \eta \mid$ $<$ 2, we find

\[ A_{FB}(B^\pm)= [ -0.24 \pm 0.42 (stat) \pm 0.19 (syst)] \% \]

limited by the  statistical uncertainty and consistent with no significant $FB$ asymmetry. The $\pm$ 0.19\% systematic uncertainty is dominated by variations due to alternative boosted decision trees and cuts.

\begin{figure}[H]
\centering
\includegraphics[height=2.8in]{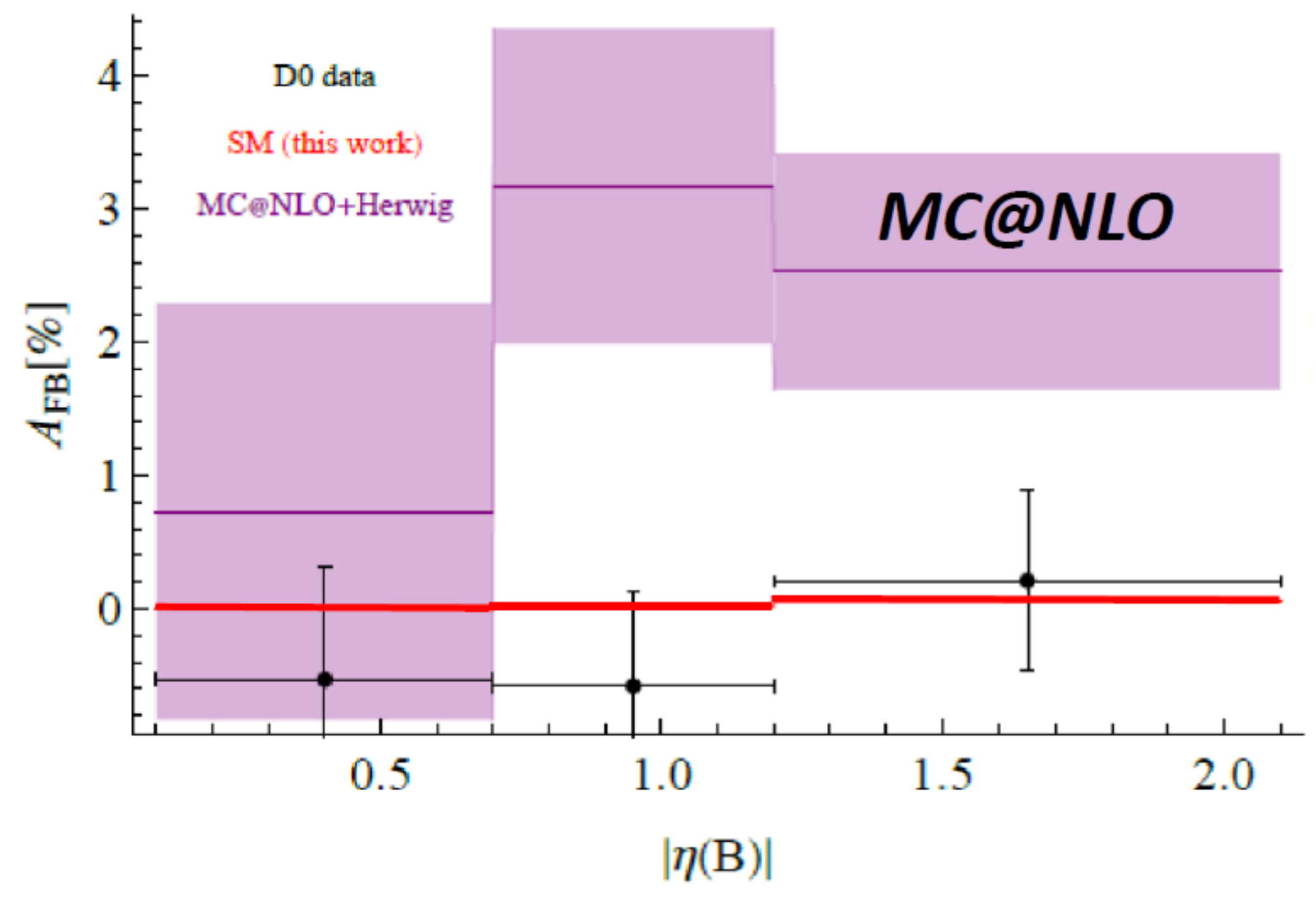}
\caption{Comparison of measured $A_{FB}(B^\pm)$ with predictions of MC@NLO \cite{MC@NLO} simulation (band) and recent calculation by C. Murphy \cite{C_Murphy} (red line). }
\label{fig:MC_NLO}
\end{figure}

An MC@NLO \cite{MC@NLO} plus HERWIG \cite{HERWIG} simulation predicts 
$A_{FB}^{MC@NLO}(B^\pm) = [ -2.31 \pm 0.34 (stat) \pm 0.44 (syst)]$ \% 
for the same  0.1 $<$ $\mid \eta \mid$ $<$ 2 range, which differs significantly from zero asymmetry and is systematically higher than the D0 measurement for all values of $\mid \eta \mid$. An improved analytic calculation by C. Murphy of Pisa \cite{C_Murphy} gives $A_{FB}(B^\pm)$ = [0.021 $\pm$ 0.008]\% for the same $\eta$ range, while reproducing the much larger observed $A_{FB}(t\overline{t})$.  These measurements, simulations, and calculations are compared in Figure~\ref{fig:MC_NLO} as a function of $\mid \eta \mid$.

\section{$A_{FB}(\Lambda_b^0)$ \cite{AFB_LambdaB}} 


The decays $\Lambda_b^0 \rightarrow J/\psi\Lambda_s^0$ with subsequent decay $J/\psi \rightarrow \mu^+ \mu^-$ and $\Lambda_s^0 
\rightarrow p \pi^-$
(and  the corresponding antiparticles) required a transverse separation of the $\Lambda$ decay and primary interaction vertices of 0.5--25 cm, and a significance of the transverse separation of the $\Lambda_b^0$ decay vertex from the primary vertex of $L_{xy}/\sigma_{L_{xy}} > $ 3.
The average transverse momentum is $<p_T(\Lambda_b^0)>$ = 9.9 GeV.  Since D0 does not have hadron identification, the 
higher momentum particle for the $\Lambda_s^0$ or $\overline{\Lambda}_s^0$ candidate was assigned the mass of the proton.  

\begin{figure}[H]
\centering
\includegraphics[height=2.25in]{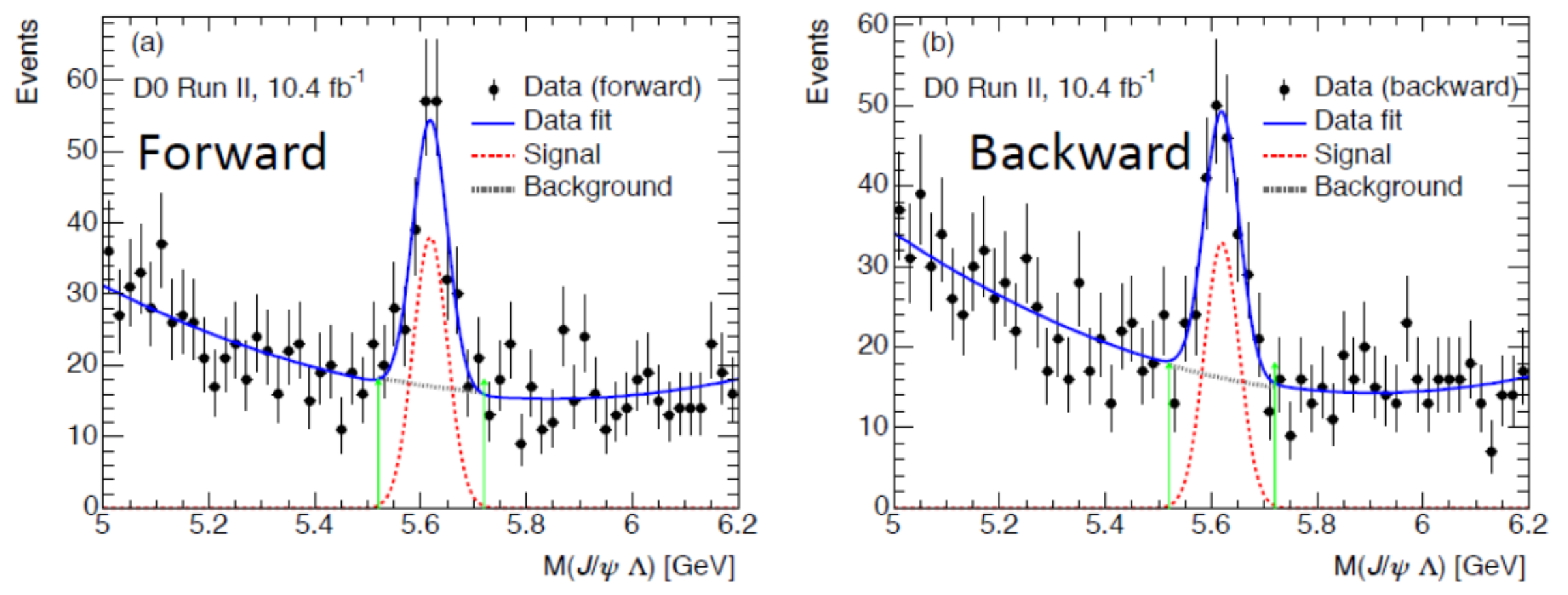}
\caption{Mass distributions for the range $ 0.5 < \mid y \mid < 1.0$ for Forward and Backward $\Lambda_b^0 \rightarrow J/\psi \Lambda$ and $\overline{\Lambda}_b^0 \rightarrow J/\psi \overline{\Lambda}$ candidates.}
\label{fig:LambdaB}
\end{figure}

In Figure~\ref{fig:LambdaB}, the data were fitted with a binned maximum-likelihood method with a Gaussian signal plus a second-order Chebyshev polynomial background distribution, resulting in a total of $842 \pm 48$ $\Lambda_b^0 + \overline{\Lambda}_b^0$ events.

The asymmetry integrated over 0.1 $<$ $\mid y \mid$ $<$ 2 for $<p_T(\Lambda_b)>$ = 9.9 GeV is measured to be
$A_{FB}(\Lambda_b^0) = 0.04 \pm 0.07 (stat) \pm 0.02 (syst)$.

\begin{figure}[H]
\centering
\includegraphics[height=2.8in]{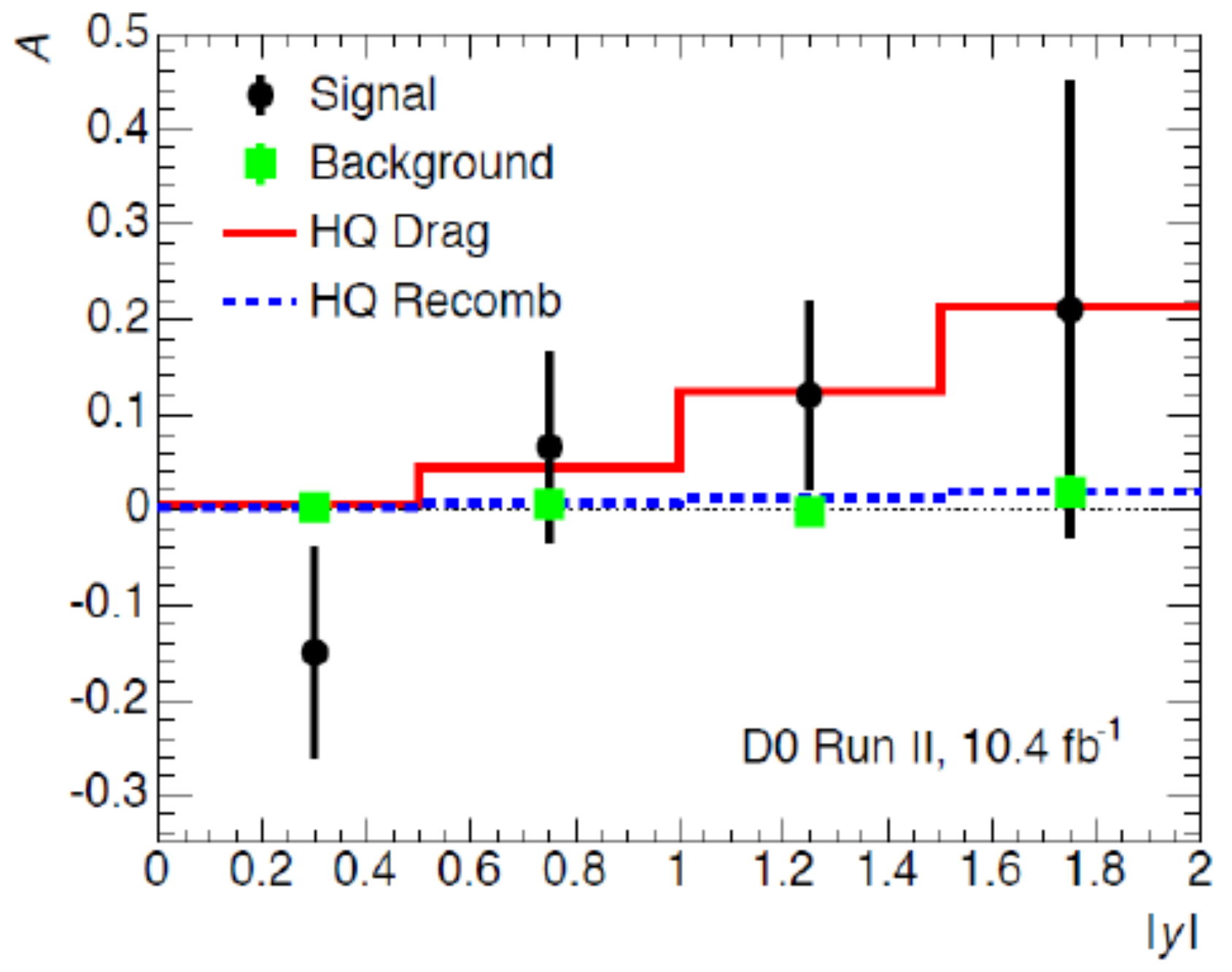} 
\caption{Forward-Backward Asymmetry for $\Lambda_b^0$ and $\overline{\Lambda}_b^0$, compared with prediction 
of the Rosner String Drag \cite{Rosner} and the Heavy Quark Recombination \cite{recombo} Models.}
\label{fig:AFB_LambdaB_w_Models}
\end{figure}

\begin{figure}[H]
\centering
\includegraphics[height=1.2in]{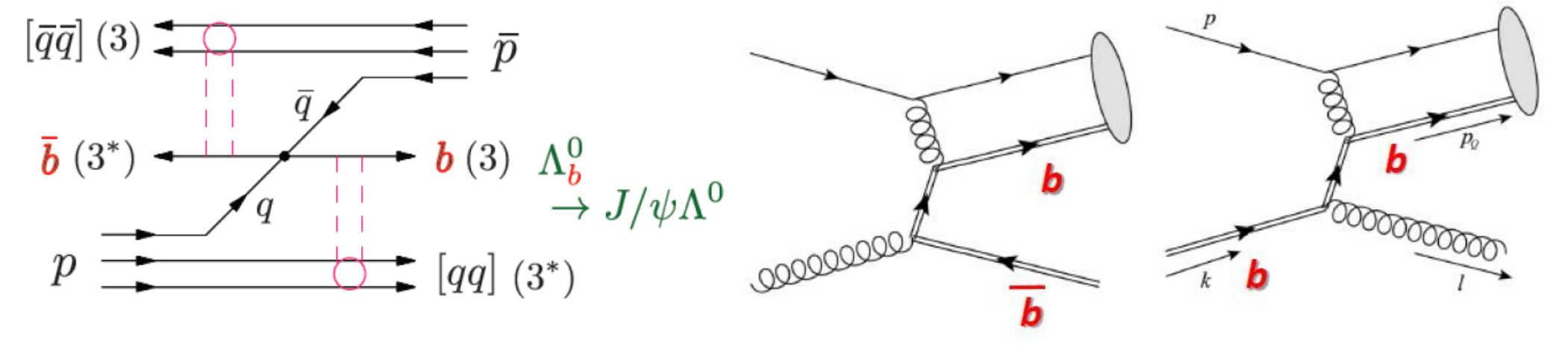}
\caption{(left) The Rosner String Drag Model \cite{Rosner} and (center and right) the Heavy Quark Recombination Model of Lai and Leibovich \cite{recombo}.}
\label{fig:Rosner_and_Recombo}
\end{figure}

As shown in Figure~\ref{fig:AFB_LambdaB_w_Models}, $A_{FB}(\Lambda_b^0)$ as a function of $\mid y \mid$ is consistent with zero but does show a tendency to increase with increasing 
$\mid y \mid$.  This is consistent with the tendency of the String Drag model by Rosner \cite{Rosner} in 
Figure~\ref{fig:Rosner_and_Recombo} where the di-quark in the proton fragment tends to predominantly pick up the produced $b$-quark giving an average increase in longitudinal momentum $<\Delta p_z>$ = + 1.4 GeV for the $\Lambda_b$ compared to \(-\) 1.4 
GeV for the $\overline{\Lambda}_b$.  In Figure~\ref{fig:Rosner_and_Recombo}, the Heavy Quark Recombination Model of Lai and Leibovich \cite{recombo} $A_{FB}(\Lambda_b^0)$ increases very little over this $\mid y \mid$ range and is consistent with $A_{FB}(\Lambda_b^0)$ = 0 and with the D0 measurement.

Both CMS and LHCb observe the ratio of $R_{pp}^{LHC} = \sigma(\overline{\Lambda}_b)/\sigma_{pp}(\Lambda_b)$ and plot this ratio as a function of the rapidity loss relative to the proton beam $y_{beam}-y_{\Lambda_b}$.  Using the expected $p\overline{p}$ charge symmetry 
$\sigma(\Lambda_b,y) = \sigma(\overline{\Lambda}_b,-y)$, it can be shown that for $p\overline{p}$ reactions that

\[ R_{pp}^{LHC} = R_{p\overline{p}} = \frac{\sigma_{p\overline{p}}(Backward)}{\sigma_{p\overline{p}}(Forward)} \]

 where

\[ R_{p\overline{p}}(\Lambda_b,\mid y \mid) = \frac{1-A_{FB}(\Lambda_b,\mid y \mid)}{ 
                       1+A_{FB}(\Lambda_b,\mid y \mid)}. \]

\begin{figure}[H]
\centering
\includegraphics[height=3.0in]{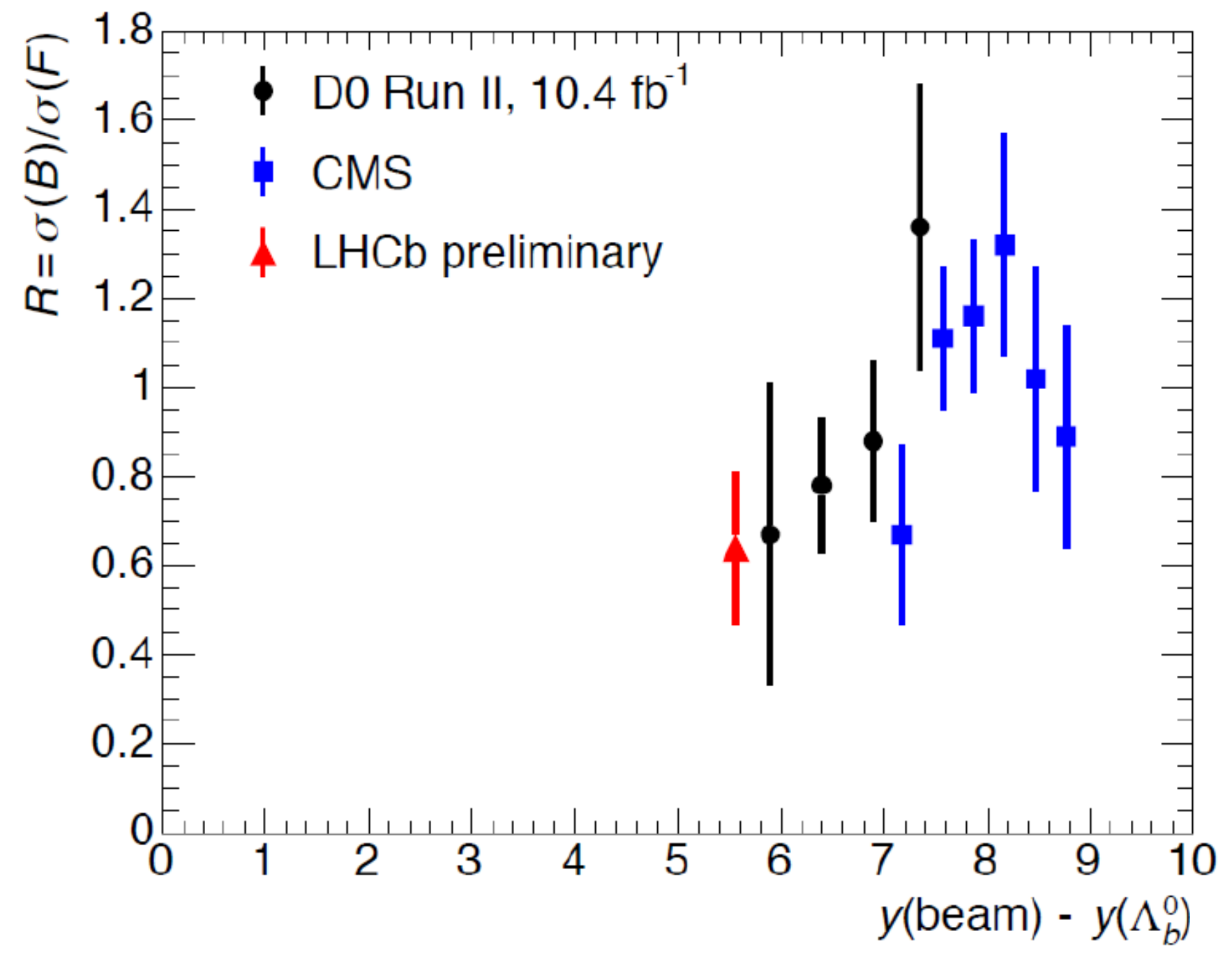}
\caption{The D0 measured values of $R_{p\overline{p}} = \sigma(B)/\sigma(F)$ for $\Lambda_b$ compared to 
$R_{pp} = \sigma(\overline{\Lambda}_b)/\sigma(\Lambda_b)$ for CMS \cite{CMS} and LHCb \cite{LHCb}.}
\label{fig:AFB_LambdaB}
\end{figure}

Thus the D0 Ratio $R = \sigma_B(\Lambda_b + \overline{\Lambda}_b)/\sigma_F(\Lambda_b + \overline{\Lambda}_b)$ can be directly plotted in Figure~\ref{fig:AFB_LambdaB} along with the CMS \cite{CMS} and LHCb \cite{LHCb} data for $R = \sigma(\overline{\Lambda}_b)/\sigma(\Lambda_b)$ showing similar behavior as a function of the rapidity loss $\Delta y = y_{beam} - y_{\Lambda_b}$.

\section{$A_{FB}(\Lambda_s^0)$ \cite{AFB_LambdaS}}   


The $\Lambda_s^0$ particles were collected under three separate trigger conditions;  1) prescaled beam crossing or minimum bias triggers (these are called the inclusive $\Lambda$ data set and consists of 2.3 Million events); 2)  events with a $\Lambda$ or $\overline{\Lambda}$ along with a $J/\psi$ which decays to $\mu^+ \mu^-$ (0.7 Million events); and 3) and events with a $\Lambda$ or $\overline{\Lambda}$ along with either a single $\mu^+$ or $\mu^-$, regardless of the charge and $particle/\overline{particle}$ correlation between the $\mu^\pm$ and the $\Lambda$ or $\overline{\Lambda}$ (53 Million events).  The D0 primary two-muon trigger resulted in the very high statistics for the $J/\psi \Lambda$ sample.  We have not fully understood the correlations between the charge of the 
single $\mu^\pm $ and the $\Lambda$ or the $\overline{\Lambda}$, so we cannot yet describe the full physics import of the single $\mu \Lambda$ events.

\begin{figure}[H]
\centering
\includegraphics[height=2.75in]{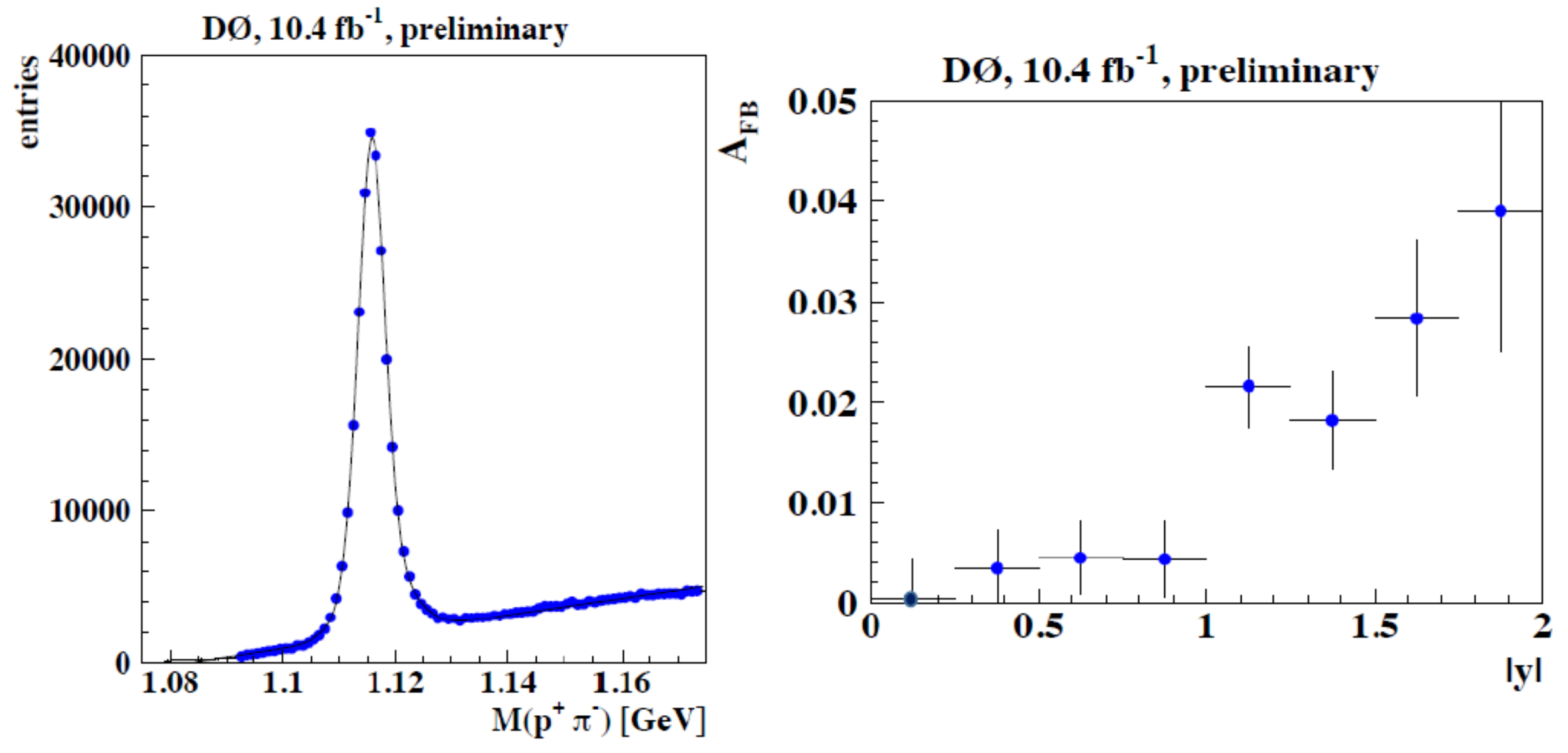}
\caption{(left) An example of a representative Mass spectrum for $\Lambda \rightarrow p \pi^+$ and $\overline{\Lambda} \rightarrow p \pi^-$ candidates. (right) Measured $A_{FB}$ as a function of $\mid y \mid$ for inclusive $\Lambda$ and $\overline{\Lambda}$.}
\label{fig:AFB_LambdaS}
\end{figure}

The plain, old time $\Lambda_s^0$ was reconstructed via the decay $\Lambda_s^0 \rightarrow p \pi^-$ and $\overline{\Lambda}_s^0 \rightarrow \overline{p} \pi^+$ as shown in Figure~\ref{fig:AFB_LambdaS}.  Again, since D0 has no particle identification for baryons, the particle with the higher momentum in the $V^0$ decay is assigned the proton mass.  The $\Lambda$ candidates were required to have 
$2<p_T(\Lambda)<25$ GeV.  

D0 has not submitted the $\Lambda$ and $\overline{\Lambda}$ analyses for publication, so these $A_{FB}(\Lambda)$ results must be considered preliminary.

\begin{figure}[H]
\centering
\includegraphics[height=3.0in]{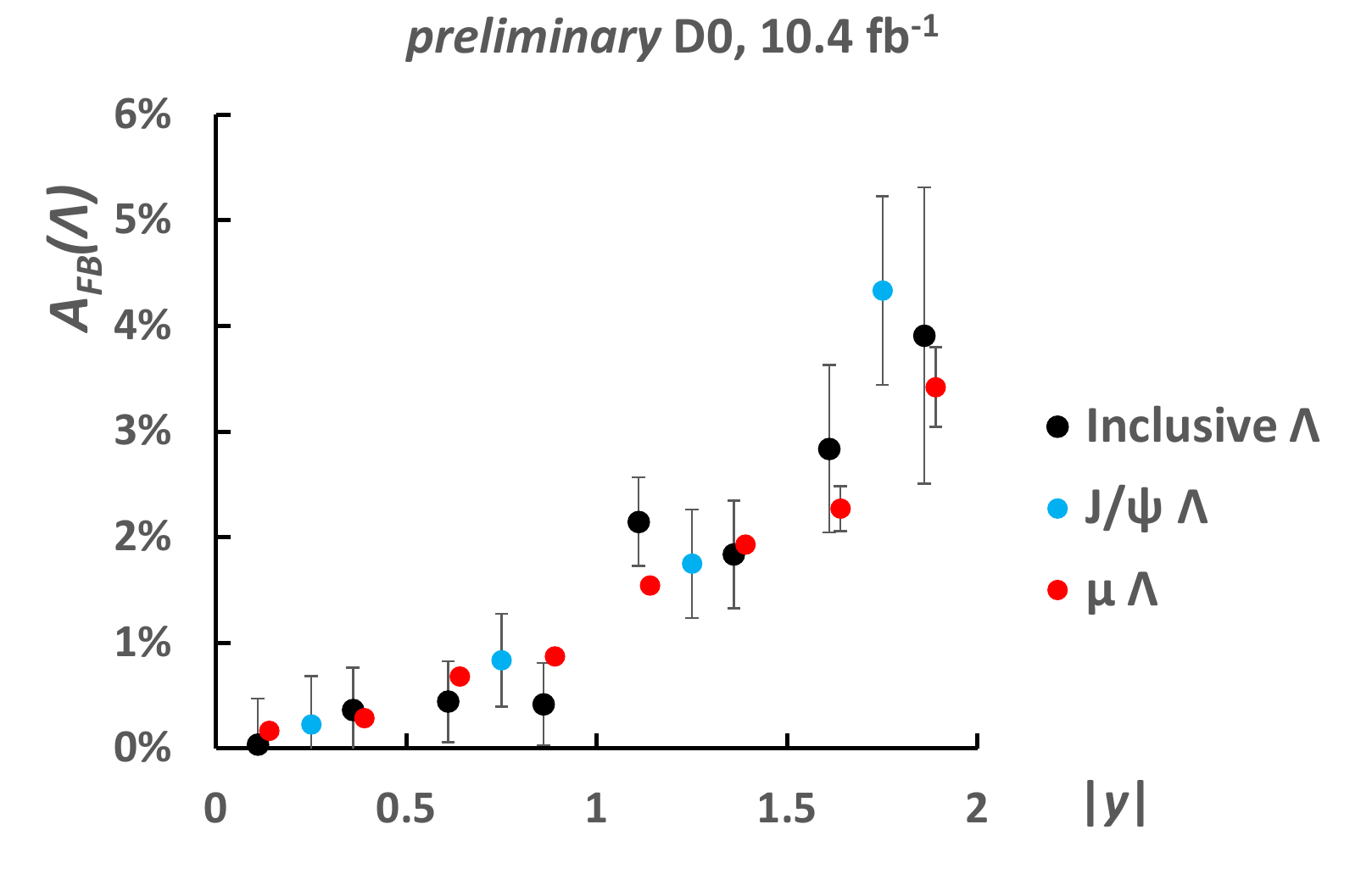}
\caption{$A_{FB}(\Lambda)$ as a function of $\mid y \mid$ for inclusive $\Lambda$,
$J/\psi \Lambda$, and $\mu \Lambda$.}
\label{fig:all_AFB_LambdaS}
\end{figure}

\begin{figure}[H]
\centering
\includegraphics[height=4.8in]{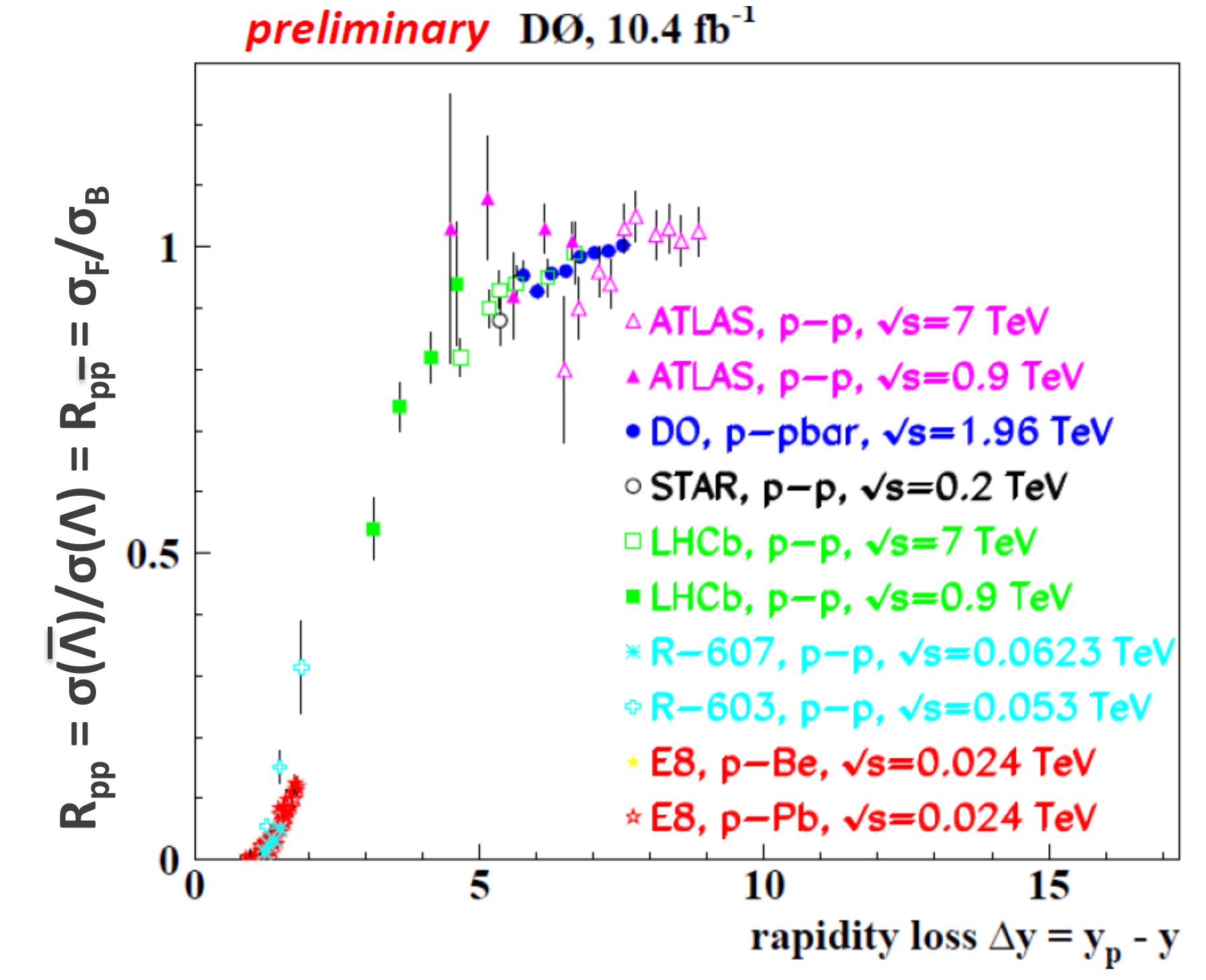}
\caption{Comparison of the D0 measured $R_{p\overline{p}} = \sigma(B)/\sigma(F)$ and
$R_{pp} = \sigma(\overline{\Lambda})/\sigma(\Lambda)$ for LHC and ISR experiments and
$R_{pA}$ for Fixed Target experiments as a function of rapidity loss $\Delta y = y_{beam} - y_{\Lambda}$:  
ATLAS \cite{ATLAS}, STAR \cite{STAR}, LHCb \cite{LHCb-Lambda}, R-607 \cite{R-607}, R-603 \cite{R-603}, 
and Fermilab E-8 \cite{E-8}.}
\label{fig:compare_R}
\end{figure}

For the inclusive $\Lambda$ and $\overline{\Lambda}$ sample, in Figure~\ref{fig:AFB_LambdaS}, $A_{FB}(\mid y \mid)$ increases with increasing $\mid y \mid$ and is statistically significantly greater than zero for 1 $<$ $\mid y \mid$.  $A_{FB}(\mid y \mid)$ for the $J/\psi \Lambda$ and for the $\mu \Lambda$ data sample also has similar behavior in Figure~\ref{fig:all_AFB_LambdaS}.  Similarly, when plotted as a function of the rapidity loss relative to the beam proton or antiproton 
$\Delta y =$ $y_{beam}-y_{\Lambda}$, 
the comparison of $\Lambda_s^0$ and $\overline{\Lambda}_s^0$  production at the $p\overline{p}$ Tevatron and for the $pp$ LHC \cite{ATLAS,STAR,LHCb-Lambda}, for the $pp$ ISR \cite{R-607,R-603}, and also for the $p-Be$ and $p-Pb$ Fixed Target \cite{E-8} data

\[ R = \frac{\sigma_{pp}(\overline{\Lambda},y)}{\sigma_{pp}(\Lambda,y)} = 
	\frac{1 - A_{FB}^{p\overline{p}}(\Lambda,y)}{1 + A_{FB}^{p\overline{p}}(\Lambda,y)} \]

exhibit very similar behavior in Figure~\ref{fig:compare_R}, which would hint at a common universal function linking the central production region near $y$ = 0 which is dominated by hard production of quark-antiquark and particle-antiparticle pairs and the forward beam fragmentation regions at high Feynman-$x$ = $p_{particle}/p_{beam}$.

\section{Conclusions and Things to Remember about the D0 Studies of $A_{FB}$}  

$A_{FB}(t\overline{t})$ is approximately +13\% over the same rapidity range.  The early difference between theory and observation motivated calculations of the Standard Model to Next-to-Next-to-Leading Oder (NNLO) to produce agreement with the data.

$A_{FB}(B^\pm)$ is consistent with zero.  The MC@NLO simulation predicts an asymmetry of approximately +2\%, while the recent calculation of Christopher Murphy gives approximately zero asymmetry.

$A_{FB}(\Lambda_b^0)$ is also consistent with zero.  More statistics are required to differentiate between the rapidity dependence of the String Drag or the Heavy Quark Recombination models.  The D0 measurement of $R(\Lambda_b^0)$ is consistent with the $R(\Delta y)$ dependence for CMS and LHCb as a function of the rapidity loss $\Delta y = y_{beam} - y_\Lambda$.

$A_{FB}(\Lambda_s^0)$ exhibits a statistically significant increase with increasing $\mid y \mid$. As the rapidity loss $\Delta y$ decreases, $R(\Lambda_s^0)$ seems to exhibit a transition from the central rapidity region at colliders to the beam fragmentation region previously studied at Fixed Target and ISR experiments.  This behavior suggests that there might be a universal curve as a function of the rapidity loss $\Delta y$.

D0 continues to produce results complementary to those of the LHC, especially those unique to $p\overline{p}$ collisions.  Stay tuned for more results!


\Acknowledgments

The D0 Collaboration thanks the staffs at Fermilab and collaborating institutions, and acknowledge 
support from the Department of Energy and National Science Foundation (United States of America); 
Alternative Energies and Atomic Energy Commission and National Center for Scientific Research/National 
Institute of Nuclear and Particle Physics (France); Ministry of Education and Science of the Russian Federation, 
National Research Center ``Kurchatov Institute'' of the Russian Federation, and Russian Foundation for Basic Research (Russia); 
National Council for the Development of Science and Technology and Carlos Chagas Filho Foundation for the Support of 
Research in the State of Rio de Janeiro (Brazil); Department of Atomic Energy and Department of 
Science and Technology (India); Administrative Department of Science, Technology and Innovation (Colombia); 
National Council of Science and Technology (Mexico); National Research Foundation of Korea (Korea); 
Foundation for Fundamental Research on Matter (The Netherlands); 
Science and Technology Facilities Council and The Royal Society (United Kingdom); 
Ministry of Education, Youth and Sports (Czech Republic); Bundesministerium f\"{u}r Bildung und Forschung (Federal Ministry 
of Education and Research) and Deutsche Forschungsgemeinschaft (German Research Foundation) (Germany); 
Science Foundation Ireland (Ireland); Swedish Research Council (Sweden);
China Academy of Sciences and National Natural Science Foundation of China (China); and 
Ministry of Education and Science of Ukraine (Ukraine).

\end{document}

%% file: econfmacros.tex
\def\beq{\begin{equation}}
\def\eeq#1{\label{#1}\end{equation}}
\def\eeqn{\end{equation}}

\def\beqa{\begin{eqnarray}}
\def\eeqa#1{\label{#1}\end{eqnarray}}
\def\eeqan{\end{eqnarray}}

\let\bar=\overbar

\def\D{{\cal D}}

\def\Dslash{\not{\hbox{\kern-4pt $D$}}}
\def\dslash{\not{\hbox{\kern-2pt $\del$}}}

\def\msb{{\bar{\ssstyle M \kern -1pt S}}}




%% file: AFB_PHG_DPF_2015-2.bbl
\begin{thebibliography}{99}
\bibitem{rapidity}
The rapidity $y$ of a particle is defined as $y$ = 0.5 * ln[$(E+p_z)/(E-p_z)$] where $E$ and $p_z$ are the energy and longitudinal momentum of that  particle.  The pseudorapidity $\eta$ of a particle is defined as 
$\eta$ = -ln (tan $\theta$/2) where $\theta$ is the polar production angle of that particle relative to the proton beam direction.
\end{thebibliography}
